\pdfoutput=1
 
\documentclass[aoas]{imsart}

\RequirePackage{natbib}
\usepackage{amsthm,amsmath,natbib}

\usepackage{amsfonts}
\usepackage{amssymb}
\usepackage{graphicx}
\usepackage{amsthm}

\usepackage{tikz}
\usetikzlibrary{arrows,shapes,trees} 
\usepackage{float}

\newcommand\independent{\protect\mathpalette{\protect\independenT}{\perp}} 
\def\independenT#1#2{\mathrel{\rlap{$#1#2$}\mkern2mu{#1#2}}} 

\newcommand\scalemath[2]{\scalebox{#1}{\mbox{\ensuremath{\displaystyle #2}}}}

\startlocaldefs
\endlocaldefs

\begin{document}

\begin{frontmatter}

\title{Structure estimation for mixed graphical models in high-dimensional data}
\runtitle{Structure Estimation for Mixed Graphical Models}

\author{\fnms{Jonas M. B.} \snm{Haslbeck}\ead[label=e1]{jonashaslbeck@gmail.com}}
\address{\printead{e1}}
\affiliation{Utrecht University}
\and
\author{\fnms{Lourens J.} \snm{Waldorp}\ead[label=e2]{waldorp@uva.nl}}
\address{\printead{e2}}
\affiliation{University of Amsterdam}

\runauthor{Haslbeck J.M.B and Waldorp L.J.}

\begin{abstract}
Undirected graphical models are a key component in the analysis of complex observational data in a large variety of disciplines. In many of these applications one is interested in estimating the undirected graphical model underlying a distribution over variables with different domains. Despite the pervasive need for such an estimation method, to date there is no such method that models all variables on their proper domain. We close this methodological gap by combining a new class of mixed graphical models with a structure estimation approach based on generalized covariance matrices. We report the performance of our methods using simulations, illustrate the method with a dataset on Autism Spectrum Disorder (ASD) and provide an implementation as an R-package.

\end{abstract}


\end{frontmatter}

\section{Introduction}
Determining conditional independence relationships through undirected graphical models, also known as Markov random fields (MRFs), is a key component of the statistical analysis of complex observational data in a wide variety of domains such as statistical physics, image analysis, medicine and more recently psychology \citep{Borsboom_Network_2013}. In many of these applications one is interested in estimating the MRF underlying a joint distribution over \textit{variables with different domains}.

As an example, consider a dataset of questionnaire responses of individuals diagnosed with Autism Spectrum Disorder (ASD), covering demographics, social environment, diagnostic measurements and aspects of well-being. A central research question in the study of Autism is how to explain individual differences in well-being. Many studies tried to answer this question by focusing on the relation between well-being and specific areas such as social functioning, cognitive ability, education and working conditions \citep[e.g.][]{magiati_cognitive_2014, anderson_predicting_2014}. While this approach provides valuable insights, one necessarily misses the full picture of all variables and might misinterpret relationships that change when additional variables are taken into account. An alternative is an integrated analysis, in which one determines the conditional independence relationships of \textit{all variables} by estimating the MRF underlying the multivariate distribution over all variables. In most datasets, however, this means that we need a method to estimate an MRF underlying a multivariate distribution over variables of different domains. In our example dataset, for instance, we have variables that are continuous (age), ordinal (IQ-scores), categorical (type of housing) and count-valued (number of different medications). 

Despite the pervasive need for a method to estimate a MRF underlying a joint distribution over mixed variables in many disciplines, so far such a general method is not available. In the present paper we address this methodological gap by combining a new class of \textit{mixed joint distributions} (see Section \ref{mixedMRF}) with a structure estimation approach based on \textit{generalized covariance matrices} (see Section \ref{gencovmat}). We thereby provide the first method that estimates a mixed MRF in which all variables are modeled on their proper domain. This avoids possible information loss due to variable transformations. In addition, our method is attractive in terms of interpretation as it is similar to the well-known Gaussian case. We introduce basic concepts in Section \ref{MRFs}, describe our estimation algorithm in detail in Section \ref{algorithm}, show performance benchmarks in Section \ref{results} and apply our method to the above dataset on ASD in Section \ref{application}. In the remainder of this section, we review existing methods to estimate MRFs underlying mixed distributions. 

Graphical models associated with distributions over one type of variable are well-known and used in many applications. The most prominent example is the Gaussian MRF underlying the multivariate Gaussian distribution. A corollary of the Hammersley-Clifford theorem \citep{Lauritzen_graphical_1996} states that zeros in the inverse covariance matrix of the multivariate Gaussian distribution indicate absent edges in the corresponding graphical model. Two classes of efficient algorithms leverage this relationship in order to estimate the Gaussian MRF: \textit{global} methods estimate the whole graph by directly estimating the inverse covariance matrix using a penalized likelihood \citep{Yuan_Model_2007, Banerjee_model_2008,friedman_sparse_2008} and \textit{nodewise} methods estimate the neighborhood of each node separately by solving a collection of sparse regression problems using the Lasso \citep{meinshausen_high-dimensional_2006}. Another graphical model that is widely used is the MRF underlying the Potts model \citep[see e.g.][]{wainwright_graphical_2008}. For the Ising model (Potts model with $m=2$ categories), \cite{Ravikumar_high-dimensional_2010} proposed a nodewise estimation method based on $\ell_{1}$-regularized neighborhood regression. This method was further extended by \cite{van_borkulo_new_2014}, who select the regularization parameter using the Extended Bayesian Information Criterion (EBIC), which is known to perform well in selecting sparse graphs \citep{foygel_high-dimensional_2014}. For general multivariate discrete distributions, \cite{loh_structure_2013} proposed an approach based on the estimation of the inverses of generalized covariance matrices.

Work on mixed graphical models is rather recent and many of the existing methods are based on non-parametric extensions of the graphical models described above: the non-paranormal \citep{liu_nonparanormal:_2009, lafferty_sparse_2012} method uses transforms that Gaussianize the data and then fits a Gaussian MRF and thereby offers a graphical model consisting of different distributions with continuous domain. Similarly, the copula-based method of \cite{Dobra_copula_2011} uses thresholded latent Gaussian variables to model ordinal variables. Other methods use non-parametric approximations such as rank-based estimators to the correlation matrix, and then fit a Gaussian MRF \citep{Xue_regularized_2012, liu_high-dimensional_2012}. 

Another way of modeling mixed graphical models is to sidestep multivariate densities altogether and relate a set of multivariate response variables of one type to multivariate covariate variables of another type. This can be done by using multiple regression or multi-task learning models \citep{Evgeniou_Multi-task_2007}. More recent approaches also allow these multiple regression models to associate covariates with mixed types of responses \citep{Yang_Heterogeneous_2009}. Also, in many machine learning learning procedures, mixed types of variables are accounted for implicitly by using suitable distance- or entropy-based measures \citep{Hastie_Elements_2009, Hsu_Hierarchical_2007}. However, the sample complexity of non-parametric methods is typically inferior to those that learn parametric models, especially in high-dimensional settings.

Parametric approaches involve methods based on latent variables that permit mixed continuous and count variables \citep{Sammel_latent_1997} or dependencies between exponential family members through a latent Gaussian MRF \citep{Rue_Approximate_2009}. While these approaches provide statistical models for mixed data, they model dependencies between observed variables using a latent layer of unobserved variables; this typically renders techniques computationally expensive and possibly intractable when estimating these models with strong statistical guarantees. A classic model for mixed data \textit{without} latent variables is the conditional Gaussian model, that combines categorical and Gaussian variables \citep{Lauritzen_graphical_1996}. Here, Gaussian variables are modeled as a multivariate Gaussian that is conditioned on all possible states of the categorical variable, which renders the model computationally intractable. The model becomes more feasible when it is restricted to pairwise or three-way interactions \citep[][respectively]{lee_learning_2012, cheng_high-dimensional_2013}; however, the general problem remains. Finally, \cite{yang_mixed_2014} introduces a way to combine different conditional distributions to their joint distribution, given that each conditional is an exponential family member.

The key idea of our paper is to generalize the \textit{generalized covariance approach} proposed by \cite{loh_structure_2013} for discrete random variables to the class of \textit{mixed joint distributions} over random variables from the exponential family introduced by \cite{yang_mixed_2014}. We thereby obtain an estimation algorithm for mixed graphical models that both models all variables on their proper domain and is easy to interpret as the estimation method is similar to the well-known Gaussian case.

\section{Estimation of mixed Markov Random Fields in high-dimensional data}\label{estimatemMRF}

\subsection{Markov random fields}\label{MRFs}


Undirected graphical models or Markov random fields (MRFs) are families of probability distributions that respect the structure of an undirected graph. An undirected graph $G = (V,E)$ consists of a collection of nodes $V = \{1, 2, \dots, p\}$ and a collection of edges $E \subseteq V \times V$. A \textit{node cutset} is a subset $U$ of nodes that breaks the graph into two or more nonempty components when it is removed from the graph. A \textit{clique} is a subset $C \subseteq V$ such that $(s,t) \in E$ for all $s,t \in C$ where $s \neq t$. A \textit{maximal} clique is a clique that is not properly contained within any other clique. The neighborhood $N(s)$ of node $s$ is defined as the set of nodes that are connected to $s$ by an edge, $N(s) := \{t \in V | (s,t) \in E \} $. The degree of a node $s$ is denoted by $deg(s) = |N(s)|$. Throughout the paper we use the shorthand $X_{\setminus s}$ for $X_{V \setminus \{s\}}$.


To each vertex $s$ in graph $G$ we associate a random variable $X_{s}$ taking values in a space $\mathcal{X}$. For any subset $A \subseteq V$, we use the shorthand $X_{A} := \{X_{s}, s \in A\}$. For three subsets of nodes, $A$, $B$, and $U$, we write $X_{A} \independent X_{B} | X_{U}$ to indicate that the random vector $X_{A}$ is independent of $X_{B}$ when conditioning on $X_{U}$. Markov random fields can be defined in terms of the global Markov property:


\newtheorem{mrf1}{Definition}
\begin{mrf1}
(Global Markov property). If $X_{A} \independent X_{B} | X_{U}$ whenever $U$ is a vertex cutset that breaks the graph into disjoint subsets $A$ and $B$, then the random vector $X := ( X_{1}, \dots , X_{p})$ is Markov with respect to the graph $G$.
\end{mrf1}

Note that the neighborhood set $N(s)$ is always a vertex cutset for $A=\{s\}$ and $B= V \setminus \{s \cup N(s)\}$. 


By the Hammersley-Clifford Theorem \citep[e.g.][]{Lauritzen_graphical_1996}, for strictly positive probability distributions, the global Markov property is equivalent to the Markov factorization property. Consider for each clique $C \in \mathcal{C}$ a clique-compatibility function $\psi_{C}(X_{C})$ that maps configurations $x_C = \{x_s, s \in V \} $ to $\mathbb{R^+}$ such that $\psi_{C}$ only depends on the variables $X_{C}$ corresponding to the clique $C$.

\newtheorem{mrf2}{Definition}
\begin{mrf1}
(Markov factorization property). The distribution of $X$ factorizes according to $G$ if it can be represented as a product of clique functions

\begin{equation}\label{eq:fac}
P(X) \propto 
\prod_{C \in \mathcal{C}} \psi_{C}(X_{C})
.
\end{equation}
\end{mrf1}

Because we focus on strictly positive distributions, we can represent (\ref{eq:fac}) in terms of an exponential family associated with the cliques $C$ in $G$

\begin{equation}\label{eq:expfam}
P(X) = 
\exp \left \{ \sum_{C \in \mathcal{C}} \theta_{C} \phi_{C}(X_{C}) - \Phi(\theta)
\right \},
\end{equation}

\noindent
where the functions $\phi_{C}(X_{C}) = \log \psi_{C}(X_{C})$ are sufficient statistic functions specified by the exponential family member at hand,  $\theta_{C}$ are parameters associated with these functions and $\Phi(\theta)$ is the log-normalizing constant

\begin{equation*}
\Phi(\theta) = \log \int_\mathcal{X}  \sum_{C \in \mathcal{C}} \theta_{C} \phi_{C}(X_{C}) \nu (dx).
\end{equation*}

\subsection{Mixed joint distributions}\label{mixedMRF}

\cite{yang_mixed_2014} introduced a special case of the form (\ref{eq:expfam}) which allows to model any combination of conditional univariate members of the exponential family within one joint distribution which respects the neighborhoods of the conditional distributions. We first describe this class of models and then provide an example.

Consider a $p$-dimensional random vector $X=(X_{1}, \dots,  X_{p})$ with each variable $X_{s}$ taking values in a potentially different set $\mathcal{X}_{s}$ and let $G = (V,E)$ be an undirected graph over $p$ nodes corresponding to the $p$ variables. Now suppose the node-conditional distribution of node $X_{s}$ conditioned on all other variables $X_{\setminus s}$ is given by an arbitrary univariate exponential family distribution

\begin{equation}\label{yang:mixed_gcond}
P(X_{s}|X_{\setminus s}) = \exp \left \{ 
E_{s}(X_{\setminus s})
 \phi_{s}(X_{s}) + C_{s}(X_{s}) - \Phi (X_{\setminus s})  
 \right \},
\end{equation}

\noindent
where the functions of the sufficient statistic $\phi_{s}(\cdot)$ and the base measure $C_{s}(\cdot)$ are specified by the choice of exponential family and the canonical parameter $E_{s}(X_{\setminus s})$ is a function of all variables except $X_{s}$.

These node-conditional distributions are consistent with a joint MRF distribution over the random vector $X$ as in (\ref{eq:fac}), that is, Markov with respect to graph $G=(V,E)$ with clique-set $\mathcal{C}$ of size at most $k$, if and only if the canonical parameters $\{E_{s} (\cdot)\}_{s\in V}$ are a linear combination of $k$-th order products of univariate sufficient statistic functions $\{\phi(X_{t}) \}_{t \in N(s)}$

\begin{equation*}\label{eq:nodecond.satisfy}
\theta_{s} + \sum_{t \in N(s)} \theta_{st} \phi_{t}(X_{t}) + ... + \sum_{t_{1}, ..., t_{k-1} \in N(s)} \theta_{t_{1}, ..., t_{k-1}} \prod_{j=1}^{k-1} \phi_{t_{j}} (X_{t_{j}}),
\end{equation*}

\noindent
where $\theta_{s\cdot} := \{\theta_{s}, \theta_{st}, ..., \theta_{s t_{2}...t_{k}}   \}$ is a set of parameters and $N(s)$ is the set of neighbors of node $s$ according to graph $G$. The corresponding joint distribution has the form

\begin{equation}\label{eq:mixed.joint.full}
\begin{split}
P(X;\theta) = \exp \left \{ 
\sum_{s \in V} \theta_{s} \phi_{s} (X_{s}) + \sum_{s \in V} \sum_{t \in N(s)} \theta_{st} \phi_{s}(X_{s}) \phi_{t}(X_{t}) + \right. \\ \left. 
\dots +  \sum_{t_{1}, ..., t_{k} \in \mathcal{C}} \theta_{t_{1}, ..., t_{k}} \prod_{j=1}^{k} \phi_{t_{j}} (X_{t_{j}}) + \sum_{s \in V} C_{s}(X_{s}) - \Phi(\theta) \right \},
\end{split}
\end{equation}

\noindent
where $\Phi(\theta)$ is the log-normalization constant. 

A limitation of this class of models is that in case they include two univariate distributions with infinite domain, they are not normalizable if \textit{neither} both distributions are infinite only from one side \textit{nor} the base measures are bounded with respect to the moments of the random variables \citep[for details see][]{yang_mixed_2014}. We return to this limitation in the discussion.

As a concrete example of (\ref{eq:mixed.joint.full}), take the Ising-Gaussian model: consider a random vector $X := (Y, Z)$, where $Y = \{Y_{1}, \dots, Y_{p}\}$ are univariate Gaussian random variables, $Z = \{Z_{1}, \dots, Z_{p}\}$ are univariate Bernoulli random variables and we only consider pairwise interactions between sufficient statistics. For the univariate Gaussian distribution (with known $\sigma^2$) the sufficient statistic function is $\phi_{Y}(Y_{s}) = \frac{Y_{s}}{\sigma_{s}}$ and the base measure is $C_{Y}(Y_{s}) = - \frac{Y_{s}^2}{2\sigma_{s}^{2}}$. The Bernoulli distribution has the sufficient statistic function $\phi_{Z_{r}} = Z_{r}$ and the base measure $C_{Z}(Z_{r}) = 0$. From (\ref{eq:mixed.joint.full}) follows that this mixed distribution has the form

\begin{equation}\label{eq:mixed.joint.pair.isinggaussian}
\begin{split}
P(Y,Z) \propto \exp \left \{
\sum_{s \in V_{Y}} \frac{\theta_{s}}{\sigma_{s}} Y_{s} + 
\sum_{r \in V_{Z}} \theta_{r} Z_{r} + 
\sum_{(s,t) \in E_{Y}} \frac{\theta_{st}}{ \sigma_{s} \sigma_{t}} Y_{s} Y_{t} + \right. \\ \left.
\sum_{(r,q) \in E_{Z}} \theta_{rq} Z_{r} Z_{q} +  
\sum_{(s,r) \in E_{YZ}} \frac{\theta_{sr}}{ \sigma_{s}} Y_{s} Z_{r} - 
\sum_{s \in V_{Y}} \frac{Y_{s}^{2}}{ 2 \sigma^{2}_{s}}
         \right \}.
\end{split}
\end{equation}

\noindent
If $X_{r}$ is a Bernoulli random variable, the node-conditional has the form

\begin{equation*}\label{yang_mixed_isinggauss_cising}
\begin{split}
P(X_{r}|X_{\setminus r}) \propto \exp 
\left \{   
 \theta_{r} Z_{r} +
 \sum_{q \in N(r)_{Z}} \theta_{rq} Z_{r} Z_{q} +   
 \sum_{t \in N(r)_{Y}} \frac{\theta_{rt}}{\sigma_{t}} Z_{r} Y_{t}  
 \right \}.
 \end{split}
\end{equation*}

\noindent
Note that this form is equivalent to the node-conditional Ising model with one term added for interactions between Bernoulli and Gaussian random variables.

If $X_{s}$ is a Gaussian random variable, the node-conditional has the form

\begin{equation*}\label{yang_mixed_isinggauss_cgauss}
\begin{split}
P(X_{s}|X_{\setminus s}) \propto \exp 
\left \{
 \frac{\theta_{s}}{\sigma_{s}} Y_{s} +     
 \sum_{t \in N(s)_{Y}} \frac{\theta_{st}}{\sigma_{s}\sigma_{t}} Y_{s} Y_{t}+   
 \sum_{r \in N(s)_{Z}} \frac{\theta_{sr}}{\sigma_{s}} Y_{s} Z_{r} -
 \frac{Y_{s}^{2}}{2 \sigma_{s}^{2}} 
 \right \}.
 \end{split}
\end{equation*}

\noindent
Not, let $\sigma = 1$, factor out $Y_{s}$ and let $\mu_{s} =  \theta_{s}+  \sum_{t \in N(s)_{Y}} \theta_{st} Y_{t}+  \sum_{r \in N(s)_{Z}} \theta_{sr} Z_{r}$. Finally, when taking $\frac{\mu_{s}^2}{2}$ out of the log normalization constant, we arrive with basic algebra at the well-known form of the univariate Gaussian distribution with unit variance

\begin{equation*}
P(X_{s} | X_{\setminus s}) = \frac{1}{\sqrt{2 \pi}} \exp 
\left \{  
- \frac{(X_{s} - \mu_{s})^{2}}{2}
\right \}.
\end{equation*}

In order to estimate the Markov random field underlying these mixed joint distributions, we use generalized covariance matrices. We introduce these in the following section.

\subsection{Generalized covariance matrices of mixed joint distributions}\label{gencovmat}

Consider a random vector $\Psi = \{X_{1}, \dots, X_{p}  \}$ that is jointly Gaussian. Then the inverse $\Gamma$ of the covariance matrix $\text{cov}(\Psi)$ is graph-structured in the sense that $\Gamma_{st} = 0$ whenever $(s,t) \not \in E$ \citep[e.g.][]{Lauritzen_graphical_1996}. While this result is well-known and leveraged by many structure estimation algorithms for the Gaussian case, the relationship between inverse covariance and graph-structure is generally unknown for general multivariate distributions. 

\cite{loh_structure_2013} improved this situation by showing that the inverses of covariance matrices over discrete random vectors are also graph-structured, when augmenting the covariance matrix appropriately with higher oder interactions. In the present paper, we extend this result to the more general class of mixed joint distributions as in (\ref{eq:mixed.joint.full}). This will allow us to use the nodewise graph algorithm proposed by \cite{loh_structure_2013} also for the estimation of mixed Markov random fields underlying mixed distributions. In the remainder of this section we first introduce necessary concepts, then illustrate the method using an example and finally state all results formally.

We require the notions of \textit{triangulation}, \textit{junction trees} and \textit{block graph structure}. Triangulation can be defined in terms of chordless cycles, which are sequences of distinct nodes $\{s_{1}, \dots, s_{\ell} \}$ such that $(s_{i}, s_{i+1}) \in E$ for all $1 \leq i \leq \ell - 1 $, $(s_{\ell}, s_{1}) \in E$  and no other nodes in the cycle are connected by an edge. Given an undirected graph $G = (V,E)$, a triangulation is an augmented graph $\widetilde G = (V, \widetilde E)$ that does not contain chordless cycles with length larger than 3. 

Any triangulation $\widetilde G$ gives rise to a junction tree representation of $G$. Nodes in the junction tree are subsets of $V$ corresponding to maximal cliques in $\widetilde G$. The intersection of two adjacent cliques $C_{1}$ and $C_{2}$ is called \textit{separator set} $S=C_{1} \cap C_{2}$. Second, any junction tree must satisfy the \textit{running intersection property}, that is, for each pair $V,C$ of cliques with intersection $S$, all cliques on the path between $U$ and $V$ contain $S$. We refer the reader to \cite{koller_probabilistic_2009} and \cite{Cowell_probabilistic_2007} for a more detailed treatment of the notions of triangulation and junction trees.

Consider a random vector $\Psi(X)$ that factors according to a graph $G$ and let $\mathcal{C}$ be the set of \textit{all} cliques in $G$. We say that an inverse covariance matrix $\Gamma = (\text{cov}(\Psi(X))^{-1})$ is \textit{ block graph-structured}, if both of the following statements are true for any $A,B \in \mathcal{C}$:
\begin{enumerate}
\item If $A,B$ are not subsets of the same maximal clique, the entry $\Gamma(A,B)$ is identically zero.
\item For almost all parameters $\theta$, the entry $\Gamma(A,B)$ is nonzero whenever $A$ and $B$ belong to the same maximal clique.
\end{enumerate}

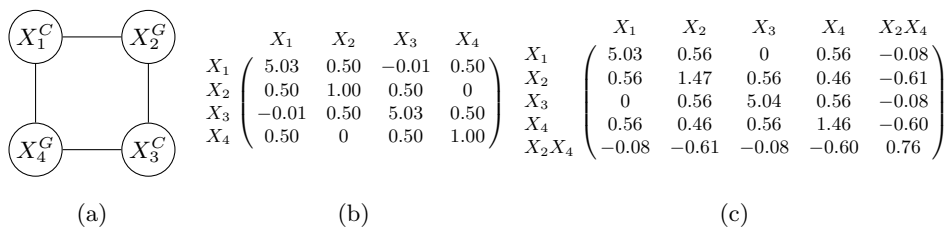
\begin{figure}[H]
  \begin{minipage}[b]{0.2\linewidth}
\centering
\begin{tikzpicture}
\tikzstyle{every node}=[draw,shape=circle];
\node[minimum size=.7cm, inner sep=1pt] (v1) at ( 90:1.5) {$X_1^C$};
\node[minimum size=.7cm, inner sep=1pt] (v2) at (45:2.121) {$X_2^G$};
\node[minimum size=.7cm, inner sep=1pt] (v3) at ( 0:1.5) {$X_3^C$};
\node[minimum size=.7cm, inner sep=1pt] (v4) at (0:0) {$X_4^G$};
\draw (v1) -- (v2)
(v2) -- (v3)
(v3) -- (v4)
(v1) -- (v4);
\end{tikzpicture}\\
\vspace{.2cm}
(a)
  \end{minipage}
  \begin{minipage}[b]{0.35\linewidth}
    \centering
    
$$
\scalemath{0.8}{
\bordermatrix{ 		  &X_{1}&X_{2}&X_{3}&X_{4}\cr
                X_{1}     & 5.03  &  0.50  & -0.01  & 0.50\cr
                X_{2}     & 0.50  &  1.00  & 0.50  & 0\cr
				X_{3}     & -0.01  &  0.50  & 5.03  & 0.50\cr
                X_{4}     & 0.50  &  0  & 0.50  & 1.00}
                }
$$
                
     	      	     \vspace{.4cm}
     	      	     (b)
  \end{minipage}
  \begin{minipage}[b]{0.45\linewidth}
      \centering
      
  $$
  \scalemath{0.8}{
  \bordermatrix{ 		  &X_{1}&X_{2}&X_{3}&X_{4} &X_{2}X_{4}\cr
                  X_{1}     & 5.03  &  0.56  & 0  & 0.56 & -0.08\cr
                  X_{2}     & 0.56  &  1.47  & 0.56  & 0.46 & -0.61\cr
  				X_{3}     & 0  &  0.56  & 5.04  & 0.56 & -0.08\cr
  				X_{4}     & 0.56  &  0.46  & 0.56  & 1.46 & -0.60\cr
                  X_{2}X_{4}     & -0.08  &  -0.61  & -0.08  & -0.60 & 0.76}
                  }
  $$
                  
       	      	     \vspace{.25cm}
       	      	     (c)
    \end{minipage}
  
\caption{(a) Ising-Gaussian Markov random field, (b) Inverse covariance matrix, (c) Inverse of augmented covariance matrix.}

\label{gencov_example}
\end{figure}

Consider an Ising-Gaussian model as in (\ref{eq:mixed.joint.pair.isinggaussian}) that factors according to the graph in Figure \ref{gencov_example} (a), where $X_1,X_3$ are Bernoulli random variables and $X_2, X_4$ are Gaussian random variables. We let all node parameters be $\theta_s = 0.1$ for all $s \in V$ and the edge parameters be $\theta_{st} = 0.5$ for all $(s,t) \in E$. Recall that if $X$ was distributed jointly Gaussian, standard theory would predict that $\Gamma$ is graph-structured. The empirically computed inverse covariance matrix $\Gamma$ of this model is displayed in Figure \ref{gencov_example} (b). While the absent edge (4,2) between the Gaussian nodes is reflected by a zero entry at $\Gamma_{4,1}$ this is not the case for the other absent edge (3,1) and $\Gamma$ is therefore not graph-structured.

Now we consider the same covariance matrix but augment it with the interaction $X_2 X_4$ and compute its inverse $\Gamma$, which is displayed in \ref{gencov_example} (c). We now have a zero entry at $\Gamma_{3,1}$, reflecting the absent edge at (3,1). However, there is a nonzero entry at $\Gamma_{4,2}$, which does not reflect the absent edge (4,2). We see that augmenting interactions to the covariance matrix renders its inverse graph structured with respect to the new, triangulated graph $\widetilde{G}$.

We will now state the general result regarding the relationship between the inverses of augmented covariance matrices and the underlying graph structure. Equipped with this result we will come back to the above example and explain the zero-pattern in \ref{gencov_example}(c).

Let $\mathcal{A} \subseteq \mathcal{C}$ be a set of cliques and define the random vector $\Psi(\phi(X);\mathcal{A}) := \{\phi_C(X_C), C \in \mathcal{A} \}$.

\newtheorem{t1}{Theorem}
\begin{t1}
\label{th_1}
Consider a mixed joint distribution $P_{\theta}(X)$ as in (\ref{eq:mixed.joint.full}) that factorizes according to a triangulated graph $\widetilde{G}$ and let $\widetilde{\mathcal{C}}$ be the set of all cliques in $\widetilde{G}$. Then the generalized covariance matrix $\emph{cov}(\Psi(\phi(X);\widetilde{\mathcal{C}}))$ is invertible, and its inverse $\widetilde{\Gamma}$ is block graph structured.
\end{t1}

The proof is based on an equality between the inverse covariance $\widetilde{\Gamma}$ and the conjugate dual $\Phi^*(\mu)$ of the log-normalizing constant $\Phi(\theta)$.  We impose the triangulation condition, because it is a sufficient condition for being able to verify the claims of Theorem \ref{th_1} in $\Phi^*(\mu)$. We provide a proof for Theorem \ref{th_1} in Section \ref{proof_t1}.

Returning to the example in Figure \ref{gencov_example}, if we take the triangulation $\widetilde{G}$ in which we augment the edge (4,2), and let $\phi(X) = X$, then the set of all cliques in $\widetilde{G}$ is equal to 

\begin{equation*}
\begin{split}
\widetilde{\mathcal{C}} = 
\left \{
X_1, X_2, X_3, X_4, 
X_1 X_2, X_2 X_3, 
X_3 X_4, 
X_4 X_1, 
X_4 X_2,  
X_1 X_2 X_4,
X_2 X_3 X_4
\right \}.
\end{split}
\end{equation*}

It can now be shown empirically that the $11 \times 11$ inverse covariance matrix $\widetilde{\Gamma} = (\text{cov}(\Psi(\phi(X);\widetilde{\mathcal{C}})))^{-1}$ reflects the graph structure of $\widetilde{G}$: there are zeros at the positions $\Gamma_{A, B}$, corresponding to the functions $\phi_A(X_A) = \prod_{s \in A} \phi_s(X_s)$  and $\phi_B(X_B) = \prod_{s \in B} \phi_s(X_s)$, whenever $A$ and $B$ are not contained in the same maximal clique. This would be the case e.g. for $A = \{1\}$ and $B = \{3\}$.


Theorem \ref{th_1} requires that we augment the covariance matrix with all cliques $C \in \mathcal{\widetilde{C}}$. A corollary of Theorem \ref{th_1} (see Corollary \ref{c_2}) states that it suffices to add all non-empty subsets of all separator sets $S$ in $\widetilde{G}$ to render $\Gamma$ graph structured. Stated formally,  let $\textbf{pow}(\mathcal{A}) = \bigcup_{C \in \mathcal{A}} \text{pow}(C)$ be the union of all $2^{|C|}-1$ nonempty subsets of all cliques $C \in \mathcal{A}$. If $\mathcal{S}$ is the set of all separator sets in $\widetilde{G}$, then we have to augment the original covariance matrix with sufficient statistics associated with the cliques in $\textbf{pow}(\mathcal{S})$ to render its inverse graph-structured.

Note that we satisfy this condition in the example in Figure \ref{gencov_example}:  We triangulate $G$ by augmenting the edge (3,1) and compute the inverse $\widetilde{\Gamma}$ of the covariance matrix over the augmented random vector $\Psi\{\phi(X); V \cup \text{pow}(\mathcal{A}) \} = \Psi\{ \phi_1(X_1), \phi_2(X_2), \phi_3(X_3), \phi_4(X_4), \phi_4(X_4)\phi_2(X_2)\}$. As we have $(3,1) \notin \widetilde{E}$ we predict $\widetilde{\Gamma}_{3,1}=0$, which is what we obtain empirically in Figure \ref{gencov_example} (c).

So far we have seen that Theorem \ref{th_1} enables us to construct a covariance matrix whose inverse reflects the graph-structure of a triangulated graph $\widetilde{G}$. However, in practice, we are not interested in the graph structure of a triangulation of $G$, but in the graph structure of the \textit{original graph} $G$. Using a neighborhood selection approach, the following corollary of Theorem \ref{th_1} enables us to estimate the \textit{original} graph $G$. Let $\mathcal{A}(s;d) := \{U \subseteq V \setminus \{s\}, |U| = d \}$, the set of all candidate neighborhoods of node $s$ with size equal to $d$.

\newtheorem{c1}{Corollary}
\begin{c1}
\label{c_1}
For any node $s \in V$ with $deg(s) \leq d$ in any graph, the inverse $\Gamma$ of the covariance matrix $\emph{cov}(\Psi(X;\{s\} \cup \emph{\textbf{pow}}(\mathcal{A}(s;d))) ) $ is graph structured such that, $\Gamma(\{s\}, B) = 0$ whenever $\{s\} \neq B \not \subseteq N(s)$. Specifically, $\Gamma(\{s\}, \{t\}) = 0 $ for all $t \notin N(s)$. 
\end{c1}

This result follows from Theorem \ref{th_1} by constructing  a particular junction tree, in which the set of candidate neighborhoods defined in Corollary \ref{c_1} separates the node $s$ from the rest of graph $G$. This new graph consisting of the node $s$ and all cliques $S(s;d)$ is triangulated, because it consists only cliques.

Recall that the candidate neighborhood is required to separate node $s$ from the rest of graph $G$. This implies that we could define a much smaller set of candidate neighborhoods if we made an additional assumption about the size of the largest clique in graph $G$ that does not include node $s$. For example, if we know that the true graph reflects the conditional independence structure of a pairwise model, it suffices to define the candidate neighborhood as  $\textbf{pow}(\mathcal{A}(s;2)))$, regardless of $deg(s)$. We return to this issue in the discussion of Algorithm \ref{alg_1} in Section \ref{algorithm}.

We can now recover the original graph $G$ by applying Corollary \ref{c_1} to each node $s \in V$ and take the row $s$ of the inverse $\Gamma$ as the support of the neighborhood $N(s)$. Because $\Gamma_{s,\setminus s}$ is a scalar multiple of the regression vector of $X_s$ upon $X_{\setminus s}$, we can use linear regression in order to estimate $\Gamma_{s,\setminus s}$. Before providing a detailed description of our nodewise method for graph-estimation in Section \ref{algorithm}, we provide the proof for Theorem \ref{th_1} and its corollaries in the following section.

\subsection{Proof of Theorem 1 and Corollary 1}\label{proof_p12}

Theorem \ref{th_1} is an extension of Theorem 1 in \cite{loh_structure_2013} in that the following theorem is about mixed Markov random fields instead of discrete Markov random fields. We first provide a proof for Theorem \ref{th_1} and then show that Corollary \ref{c_1} follows.

\subsubsection{Proof of Theorem 1}\label{proof_t1}

We follow the proof of  \cite{loh_structure_2013}, which consists of two parts. First, we have establish the equality of the inverse covariance matrix of sufficient statistics $(\text{cov}_{\theta} \{ \phi(X) \})^{-1}$ and the Hessian of the conjugate dual $\nabla^2 \Phi^{*}(\mu)$ of the log normalizing function $\Phi(\theta)$

\begin{equation}\label{lw_key_eq}
(\text{cov}_{\theta} \{ \phi(X) \})^{-1}  = \nabla^2 \Phi^{*}(\mu)
\end{equation}

\noindent
where the conjugate dual $\Phi^*$ of a function $\Phi$ is defined as

\begin{equation}\label{dual}
\Phi^{*}(\mu) := \sup_{\theta \in \Omega} \{ \langle \mu, \theta \rangle   - \Phi(\theta) \},
\end{equation}

\noindent 
where $\mu \in \mathbb{R}^d$ is a fixed vector of mean parameters of the same dimension as $\theta$, $ \langle \mu, \theta \rangle$ is the Euclidean inner product $\sum_{i=1}^m \mu_{i} \theta_{i}$ and $\mu \in \mathcal{M}$, where $ \mathcal{M} := \{ \mu \in \mathbb{R}^d \;| \; \exists \; P \; s.t.\; \mathbb{E}_{P} [ \phi_{\alpha}(X) ] = \mu \}$. We define the mean parameter $\mu_{\alpha}$ associated with a sufficient statistic $\phi_{\alpha}$ as its expectation with respect to $P_{\theta}(x)$

\begin{equation}\label{meanpar}
\mu_{\alpha} := \mathbb{E}_{P} [ \phi_{\alpha}(X) ].
\end{equation}

Note that, by assumption, $\Phi(\theta)$ is finite, which implies that $\mathcal{M}$ is bounded. For an excellent treatment on the conjugate dual of the log normalizing function we refer the reader to \cite{wainwright_graphical_2008}. 

\cite{loh_structure_2013} showed that equality (\ref{lw_key_eq}) holds for regular, minimal exponential family distributions with a finite log-normalizing constant $\Phi(\theta)$, that is, $\theta \in \Omega = \{ \theta: \Phi(\theta) < \infty \}$ and we refer the reader to their paper for the proof of this result. Using this result, for the first part of the proof it remains to be shown that these requirements hold for the class of mixed distributions as in (\ref{eq:mixed.joint.full}). \cite{yang_mixed_2014} show that the class is regular and has a finite log-normalizing constant, and minimality is established by Lemma \ref{l_2}:

\newtheorem{l2}{Lemma}
\begin{l2}
\label{l_2}
The class of mixed distributions as in \ref{eq:mixed.joint.full} is minimal.
\end{l2}

For the proof of Lemma \ref{l_2} see Appendix \ref{proof_l1}. 

The second part of the proof is to show that the graph structure claimed in Theorem \ref{th_1} holds in $\nabla^2 \Phi^{*}(\mu)$. In general there is no straight-forward way to compute $\nabla^2 \Phi^{*}(\mu)$, because we do not know how to compute derivatives of $\Phi(\theta)$ in (\ref{dual}), which depends on all cliques in the graph. However, because we require the graph to be triangulated, we can represent $P_\theta$ as a factorization of marginal distributions of cliques $C \in \mathcal{C}$ and separator sets $S \in \mathcal{S}$ \citep{Lauritzen_graphical_1996, Cowell_probabilistic_2007}

\begin{equation}\label{triang_fact}
P_{\theta} 
= 
\frac{\prod_{C \in \mathcal{C}} P_{C}(X_C)}{\prod_{S \in \mathcal{S}} P_{S}(X_S)}.
\end{equation}

If we plug the mixed joint distribution form as in (\ref{eq:mixed.joint.full}) into (\ref{triang_fact}) it is explicit that each term in the factorization depends only on the corresponding clique or separator set

\begin{equation}\label{triang_fact_yang}
P_{\theta} 
=
\frac{ 
\prod_{C \in \mathcal{C}} \exp \left \{  
\frac{
\theta_{C} 
\prod_{j=1}^{|C|} \phi_{t_{j}}(X_j)  +
\sum_{r \in C} C_r (X_r) 
}{
\log \int_{x\in\mathcal{X}^{|C|}} 
\theta_{C} 
\prod_{j=1}^{|C|} \phi_{t_{j}}(X_j)  +
\sum_{r \in C} C_r (X_r) 
}
    \right   \}
}
{ 
\prod_{S \in \mathcal{S}} \exp \left \{  
\frac{
\theta_{S} 
\prod_{j=1}^{|S|} \phi_{t_{j}}(X_j)  +
\sum_{r \in S} C_r (X_r) 
}{
\log \int_{x\in\mathcal{X}^{|S|}} 
\theta_{S} 
\prod_{j=1}^{|S|} \phi_{t_{j}}(X_j)  +
\sum_{r \in S} C_r (X_r) 
}
    \right   \}
}.
\end{equation}

We obtain $\Phi^{*}(\mu)$ of $\Phi(\theta)$ via its equality relation to the negative Shannon entropy $H(P_\theta)$ of the distribution $P_\theta$

\begin{equation}\label{triang_fact_dual}
\scalemath{0.95}{
\begin{split}
\Phi^*(\mu) &= 
- H(P_{\theta}(\mu)) = 
- \mathbb{E}[\log P_\theta] =\\
&-
 \sum_{C \in \mathcal{C}} 
 \left (
 \theta_C  \mu_{j \in C} + \sum_{r \in C} C_r (\mu_r) - \Phi_C(\theta) 
 \right )
\\ &-
 \sum_{S \in \mathcal{S}} 
 \left (
 \theta_S \mu_{j \in S} + \sum_{r \in S} C_r (\mu_r) - \Phi_S(\theta) 
 \right ).
\end{split}
}
\end{equation}

Note, that the factorization (\ref{triang_fact_yang}) turns into a sum in the conjugate dual representation (\ref{triang_fact_dual}), which enables to verify the claims of Theorem \ref{th_1} by taking partial derivatives: 

consider two subsets $A,B \in \mathcal{C}$ that are not contained in the same maximal clique $C$. As $P_{\theta}(X)$ is Markov with respect to the triangulated graph $G$, all interaction parameters $\theta_{A, \dots, B}$ associated with both $A$ and $B$ are equal to zero. When differentiating expression (\ref{triang_fact_dual}) with respect to $\mu_A$ all terms that do not involve $\mu_A$ drop. Now, when taking the second derivative with respect to $\mu_B$, all terms that do not involve $\mu_B$ drop. The only terms left are terms involving both $\mu_A$ and $\mu_B$. However, these terms are products involving a $\theta_{A, \dots, B} = 0$ and therefore we obtain $\frac{\partial^2 \Phi^*(\mu)}{\partial \mu_A \partial \mu_B} = 0$. Together with the equality (\ref{lw_key_eq}), this proves part (1) of the graph structure in Theorem \ref{th_1}.

Turning to part (2), if $A,B$ are part of the same maximal clique we have $\theta_{A, \dots, B} \not = 0$ for at least one parameter $\theta_{A, \dots, B}$. Taking partial derivatives with respect to $\mu_A$ and $\mu_B$ preserves only terms involving both $\mu_A$ and $\mu_B$. Assuming the $\theta$'s are drawn from a continuous distribution,  $\frac{\partial^2 \Phi^*(\mu)}{\partial \mu_A \partial \mu_B}$ is almost surely nonzero. Together with equality (\ref{lw_key_eq}), this proves part (2) of Theorem \ref{th_1}.

\subsubsection{Proof of Corollary 1}

We take the conjugate dual $\Phi^*(\mu)$ of $\Phi(\theta)$ corresponding to a model of the form (\ref{eq:mixed.joint.full}) including only terms for cliques $(\{s\} \cup \textbf{pow}(\mathcal{A}(s;d)))$. As above, we take derivatives with respect to $\mu_s$ and $\mu_B$ and all terms drop that do not contain both $\mu_s$ and $\mu_B$. Whenever $\{s\} \neq B \not \subseteq N(s)$, all terms including both $\mu_s$ and $\mu_B$ are equal to zero, because $P_{\theta}(X)$ is Markov with respect to $G$. Therefore, whenever $\{s\} \neq B \not \subseteq N(s)$, the partial derivative of $\Phi^*(\mu)$ with respect to  $\mu_s$ and $\mu_B$ is equal to zero. With equality (\ref{lw_key_eq}), this proves the claims of Corollary \ref{c_1}.

\subsection{Nodewise regression algorithm for structure estimation of mixed graphical models}\label{algorithm}

In this section we describe how to use neighborhood-regression together with the result in Corollary \ref{c_1} to estimate a mixed MRF from data. Recall that we obtain the whole graph $G=(V,E)$ by recovering the neighborhood $N(s)$ of all $s \in V$. Corollary \ref{c_1} enables us to construct an inverse covariance matrix $\Gamma$ that reflects the neighborhood $N(s)$. We can therefore estimate the neighborhood of $s$ by estimating the row $s$ in $\Gamma$. Recall that this row $s$ is a scalar multiple of the regression vector of $X_s$ upon on $X_{\setminus s}$ \citep[see e.g.][]{Lauritzen_graphical_1996}. 

This means that we construct a random vector $\Psi(\phi(X);\{s\} \cup \emph{\textbf{pow}}(\mathcal{A}(s;d))$ for each node $s$ as in Corollary \ref{c_1} and we then predict the node s $\Psi_s$ by all other terms  $\Psi_{\setminus s}$ in $\Psi$. In order to get a sparse parameter vector $\widehat{\theta}$ , we apply $\ell_1$-regularization, whose strength is controlled by the regularization parameter $\lambda_n$. The parameter vector $\widehat{\theta}$ is estimated by maximizing the $\ell_1$-penalized log likelihood. In case node $s$ is associated with a Gaussian random variable, this is equivalent to solving the $\ell_1$-penalized least-squares problem

\begin{equation}\label{lasso}
\widehat{\theta} = \arg \min_{||\theta||_{1} \leq b_{0} \sqrt{k}} 
\big \{
||\Psi_s - \Psi_{\setminus s} \theta||_2 + 
\lambda_{n} ||\theta||_{1}
\big \},
\end{equation}

\noindent
where $b_{0} > ||\widetilde{\theta}||_{1}$ is a constant and $k$ is the sparseness of the population parameter vector $\widetilde{\theta}$. For non-Gaussian variables a (link) function is used to define a linear relation between the dependent variable $\Psi_s$ and its predictors $\Psi_{\setminus s}$ \citep[for details see][]{Friedman_Regularization_2010}.

Solving the lasso-problem in (\ref{lasso}) for all $s \in V$ yields two estimates $\widehat{\theta}_{i,j}$ and $\widehat{\theta}_{j,i}$ for each edge. We combine these estimates with an AND-rule (both parameters have to be nonzero, otherwise the edge is absent; the value is the average) or an OR-rule (at least one parameter has to be nonzero; the parameter value is the average over the non-zero parameters).

Further, we assign the parameters $\lambda_{n}, \tau_{n}$ to the scaling

\begin{equation}\label{scaling}
\lambda_{n} \succsim  \sqrt{d} ||\widetilde{\theta}||_{2} \sqrt{\frac{\log p}{n}}, \qquad \tau_{n} \asymp \sqrt{d} ||\widetilde{\theta}||_{2} \sqrt{\frac{\log p}{n}},
\end{equation}

\noindent
where $||\widetilde{\theta}||_{2}$ is the $\ell_2$-norm of the population parameter vector $\widetilde{\theta}$ and $d$ is the degree of the true graph. By $\succsim$ and $\asymp$ we mean asymptotically larger and asymptotically over, respectively. Note the inequalities (\ref{scaling}) contain the population parameter vector $\widetilde{\theta}$. In order to compute $\tau_n$, we use the estimated parameter vector $\widehat{\theta}$ and we select $\lambda_{n}$ using 10-fold cross-validation. We thereby obtain the following nodewise algorithm for structure estimation:

\newpage

\newtheorem{a1}{Algorithm}
\begin{a1}
\label{alg_1}
(Nodewise regression method)
\begin{enumerate}
\item Construct the vector $\Psi$ according to Corollary \ref{c_1}.
\item Select a parameter $\lambda_{n}$ using 10-fold cross-validation
\item Solve the lasso regression problem (\ref{lasso}) with parameter $\lambda_{n}$, and denote the solution by $\widehat{\theta}$.
\item Threshold the entries of $\widehat{\theta}$ at level $\tau_{n}$
\item Combine the neighborhood estimates with the AND- or OR-rule and define the estimated neighborhood set $\widehat{N(s)}$ as the support of the thresholded vector
\end{enumerate}
\end{a1}

As we do not know the degree of the true graph $d$, we have to make an assumption about $d$. The straightforward choice would be $d = |V| - 1$ which puts no constraints on the true graph. However, note that the computational complexity of Algorithm \ref{alg_1} is $\mathcal{O}(2^d \log p)$, which means choosing $d = |V| - 1$ renders the algorithm unfeasible for all but small graphs.

In order to make an informed assumption about $d$, note that the choice of $d$ has two consequences in Algorithm \ref{alg_1}: first, it influences the threshold $\tau_n$ in (\ref{scaling}), which reflects the sparsity assumption necessary for asymptotic consistency. Second, $d$ determines the size of candidate neighborhoods added to the covariance matrix (see Corollary \ref{c_1}). If we now only consider satisfying the requirements of Corollary \ref{c_1}, we can choose $d$ as the size of the largest clique in graph $G$ that does not contain $s$. For example, in the case that the true model is a pairwise model, we would choose $d=2$ instead of the degree $d$ of $G$, which depends on the size and density of graph $G$. The cost of the assumption about $d$ is, with either interpretation, that we miss edges belonging to cliques with size $|C| > d$.

For the sample complexity $n \succsim d^3 \log p$, where $d$ is the maximal degree in the true graph $G$, \cite{loh_structure_2013} prove asymptotic consistency for Algorithm \ref{alg_1}. For details we refer the reader to their paper.

Because asymptotic considerations provide little information on how well a method performs in practical situations, we use simulations to test the performance of our method in recovering the graph from data in settings that resemble typical situations in exploratory data analysis.

\section{Simulations}\label{results}

We consider the following six graphical models: Binary-Gaussian, Binary-Poisson, Binary-Exponential, and Multinomial with $m = 2, 3,$ or $4$ categories. In the three mixed graphical models, half of the nodes are binary. We used the Potts model \citep[see e.g.][]{wainwright_graphical_2008} to specify interactions between categorical variables: if we have $m$ categories,  $X_{s}, X_{t}$ are two arbitrary nodes and $j,k \in m$, then $\theta_{st,jj} = 1$ and $\theta_{st,jk} = 0$, where $j \not = k$. Interactions between binary variables $X_{s} \in \{0,1\}$ and continuous variables $X_{t}$ are specified as follows: if $X_{s} = 1$ then $\theta_{st} = 1$ and if $X_{s} = 0$ then $\theta_{st} = 0$. All graphs were random graphs \citep{erdoes_random_1959} with $p = 16$ nodes and we varied sparsity by varying edge-probabilities $P_{edge} \in \{.1, .2, .3\}$. We also varied the ratio between observations and variables $\frac{n}{p} \in \exp \{0,1,2,3,4,5\}  \approx \{1, 3, 7, 20, 55, 148\}$ and the size of augmented cliques (interactions) $d \in \{1,2,3\}$, where $d=1$ means that we use the original covariance matrix. Noise was added by multiplying each edge-weight by a draw from a uniform distribution $\mathcal{U}(.3,1)$. All thresholds were set to zero, only for Poisson and Exponential nodes we set the threshold to $-.1$ in order to avoid $\lambda = 0$ in case $N(s) = \{ \emptyset \}$. All edge-weights were additionally multiplied by $-1$, which led to less extreme category-potentials and therefore more variance in the categorical variables. 

In the Binary-Gaussian case, the variance of the Gaussian univariate distributions was set to 1 and the whole random graph including noise was resampled until the weighted adjacency matrix of the Gaussian sub-graph was positive definite. For categorical variables, we required that each category is present in the data and that the category with the lowest frequency is present in more than $10 \%$ of the cases. The reason for the second condition is that we need non-zero variance in every possible response vector in 10-fold cross-validation. For the same reason, we required for Poisson variables that the category with the highest frequency is present in less than $90\%$ of the cases. In order to meet this requirement, we take the columns that do not meet the requirements and replace random row-entries by samples from categories with equal probabilities (categorical case) or by a draw from a Poisson distribution with $\lambda = \frac{1}{2}$ (Poisson case) until the requirement is met.

We combined parameters specifying the interaction between two categorical variables with the OR-rule and used the AND-rule to combine estimates between regressions.

\begin{figure}
\centering
\includegraphics[width=1\textwidth]{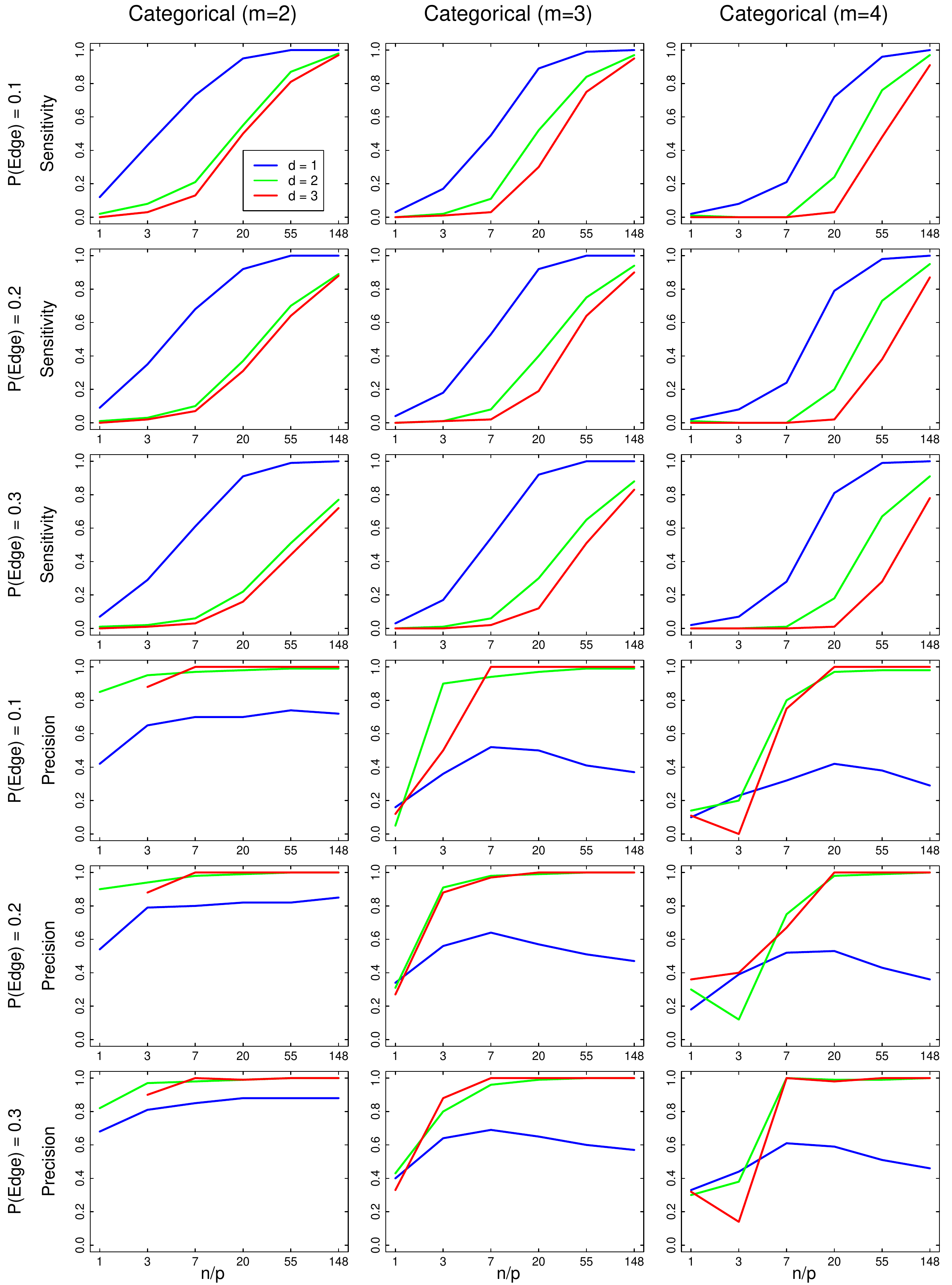}
  \caption{Simulation results: The first three rows show sensitivity, the last three rows precision of our method. We augmented cliques of size $d=\{1,2,3\}$ for different combinations of sparsity and rescaled sample size $n/p$ for graphs consisting of categorical variables with $m = 2, 3 or 4$. Missing data points in precision are due to the fact there were no edges estimated in any of the 100 iterations.}
\label{results_catcat}
\end{figure}

\begin{figure}
\centering
\includegraphics[width=1\textwidth]{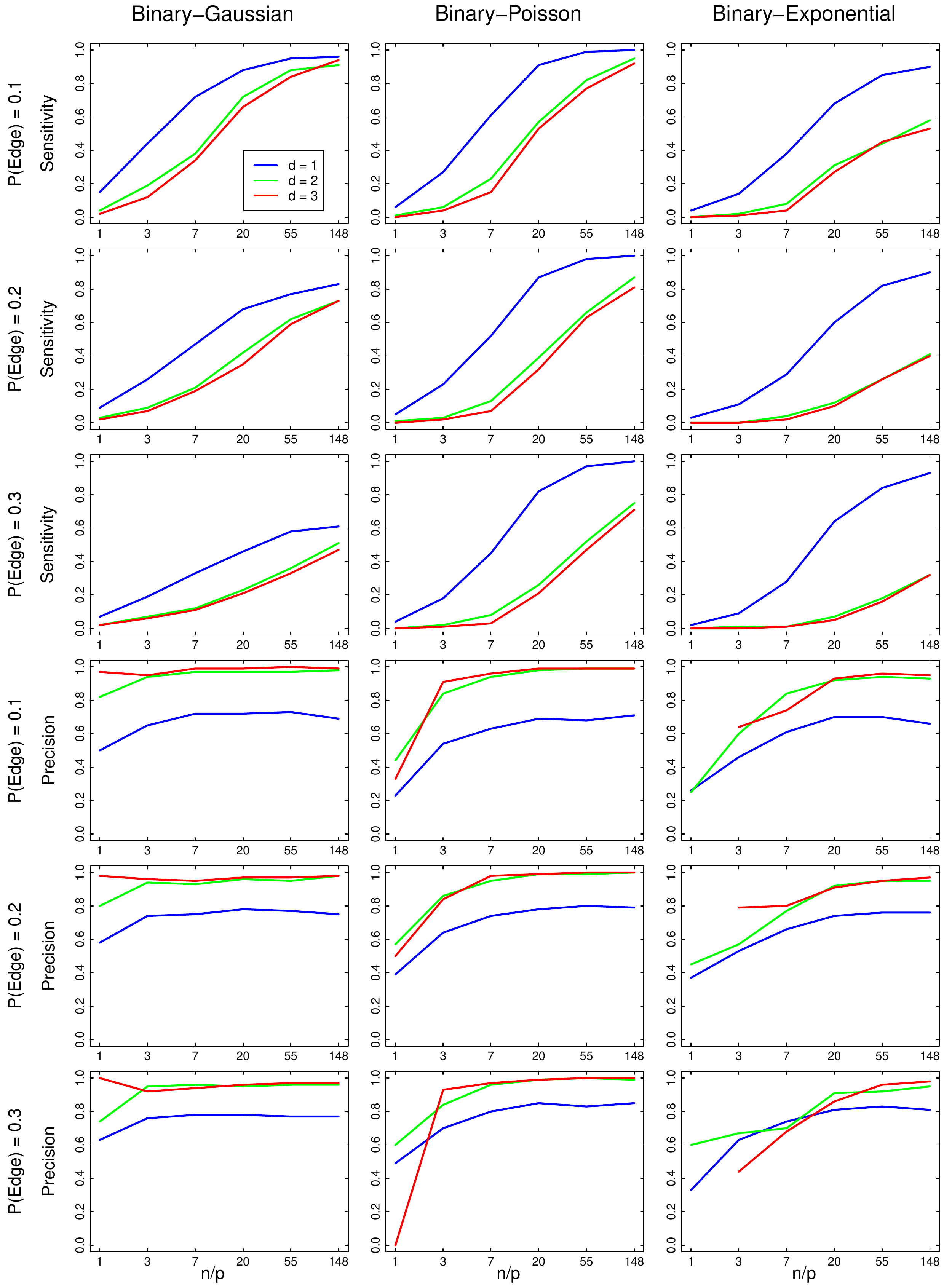}
  \caption{Simulation results: The first three rows show sensitivity, the last three rows precision of our method. We augmented cliques of size $d=\{1,2,3\}$ for different combinations of sparsity and rescaled sample size $n/p$ for Binary-Gaussian, Binary-Poisson and Binary-Exponential graphs. Missing data points in precision are due to the fact there were no edges estimated in any of the 100 iterations.}
\label{results_contcat}
\end{figure}

Sensitivity and precision of our method for different combinations of sparsity ($P_{edge}$) and $n/p$-ratio for graphs consisting of categorical random variables with $m = \{2,3,4\}$ categories is shown in Figure \ref{results_catcat}. In all conditions with $d=1$, sensitivity quickly converges to 1 with increasing observations. However, in this setting precision does not converge to 1. This is consistent with our theory, as we require in Corollary \ref{c_1} that all candidate neighborhoods with size up to larger or equal to the largest clique in the true graph are augmented. As we sampled data from a pairwise model, the largest clique in the true graph has size $d = 2$ and we violate this requirement. For $k=2,3$ we satisfy the requirements in Corollary \ref{c_1} and in these settings Algorithm \ref{alg_1} converges in precision. The general performance drops when the number of categories $m$ increases, because more parameters have to be estimated. We also observe that general performance drops with lower sparsity, however the effects are small. Note that for the binary case $m=2$ we obtain qualitatively similar results compared to \cite{loh_structure_2013} and others who used a similar estimation method \citep{Ravikumar_high-dimensional_2010, van_borkulo_new_2014}.

Note that we observe a peculiar pattern in precision in the categorical MRFs with $m>2$ categories. Precision increases initially with observations, but then \textit{decreases} again. This is a consequence of the OR-rule, which we use to combine several parameters describing the interaction between two categorical variables into one edge-parameter. As only one of the $(m-1)^2$ parameters has to be nonzero to render the corresponding edge nonzero, the algorithm becomes more liberal when the number of categories is high. While these spurious parameters are put to zero by the $\tau_n$-threshold (\ref{scaling}) for small $n$, the threshold decreases with increasing $n$ such that these spurious parameters are not set to zero anymore.

Figure \ref{results_contcat} shows sensitivity and precision of our method for the same combinations of sparsity and $n/p$-ratio as above for Binary-Gaussian, Binary-Poisson and Binary-Exponential. Similarly to the categorical case, we see in conditions with $d=1$ that sensitivity seems to converge to 1 with increasing $n$ in all conditions but the Binary-Gaussian cases with sparsity $> .1$. In conditions with $d=2,3$ sensitivity seems to converge to 1 in the Binary-Gaussian case with sparsity $P_{edge} = .1$. In all other conditions with $d=2,3$ sensitivity increases with growing $n$. Precision converges most quickly for the Binary-Gaussian case, followed by the Binary-Poisson and Binary-Exponential case. Sparsity seems to have no impact on the precision in any of the three mixed graphs. Similarly to the categorical case, precision converges to 1 despite the fact that the theoretical sample complexity is not satisfied. As above, missing data points in precision is due to the fact that the algorithm often estimates zero edges.

\section{Application to ASD Data}\label{application}

In this section we return to the exploratory data analysis problem introduced at the beginning of the paper. We estimate the MRF underlying a dataset consisting of responses to a questionnaire of 3521 individuals from the Netherlands diagnosed with Autism Spectrum Disorder (ASD). The variables cover demographics, social environment, diagnostic measurements and aspects of well-being \citep[for details see][]{Begeer_Allemaal_2013}. 

\begin{figure}
\centering
\begin{minipage}[b]{1\linewidth}
    \centering
	\includegraphics[width=1\textwidth]{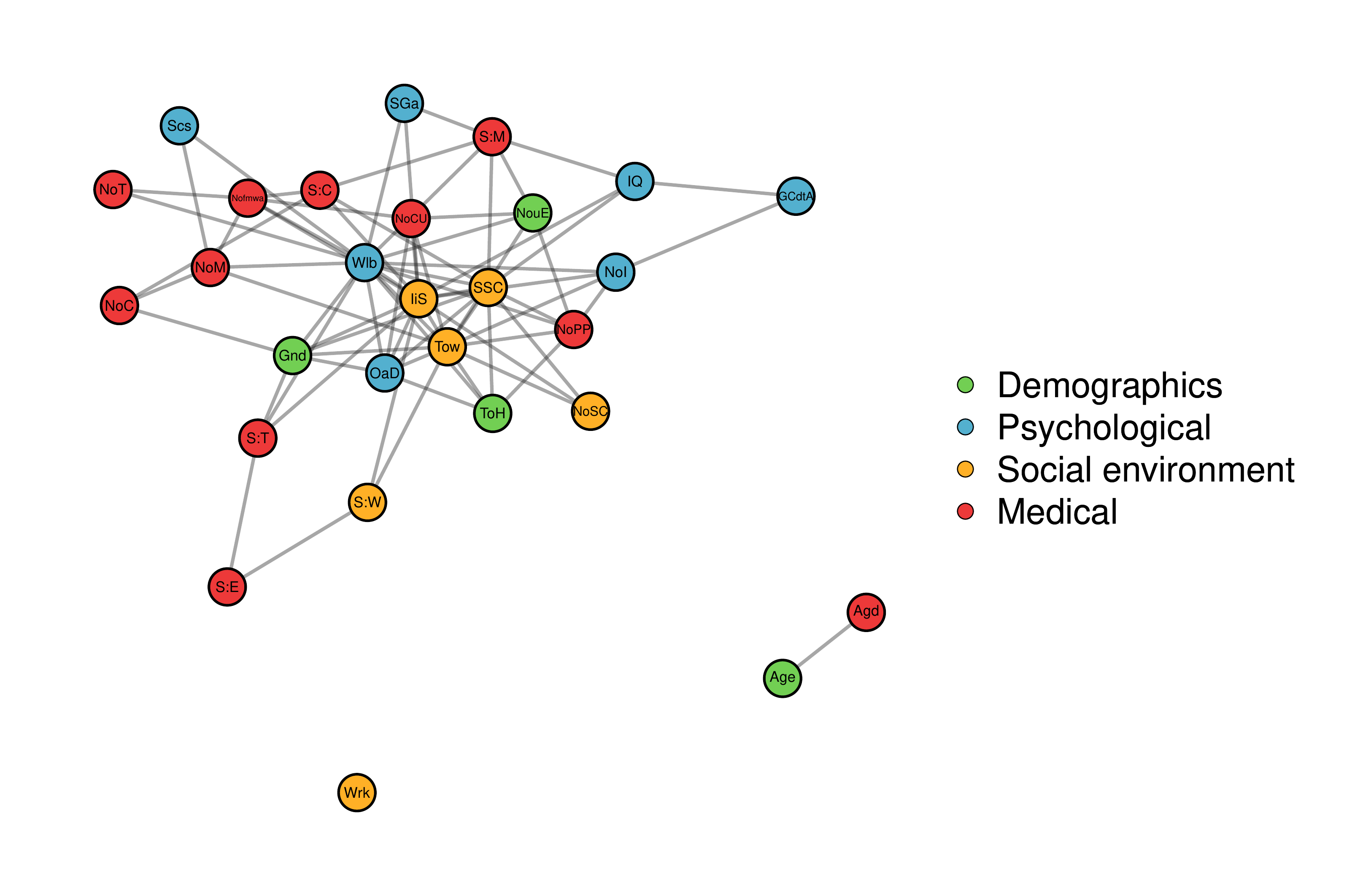}
(a) Mixed Graphical Model

  \end{minipage}
  \begin{minipage}[b]{1\linewidth}
    \centering
    \includegraphics[width=1\textwidth]{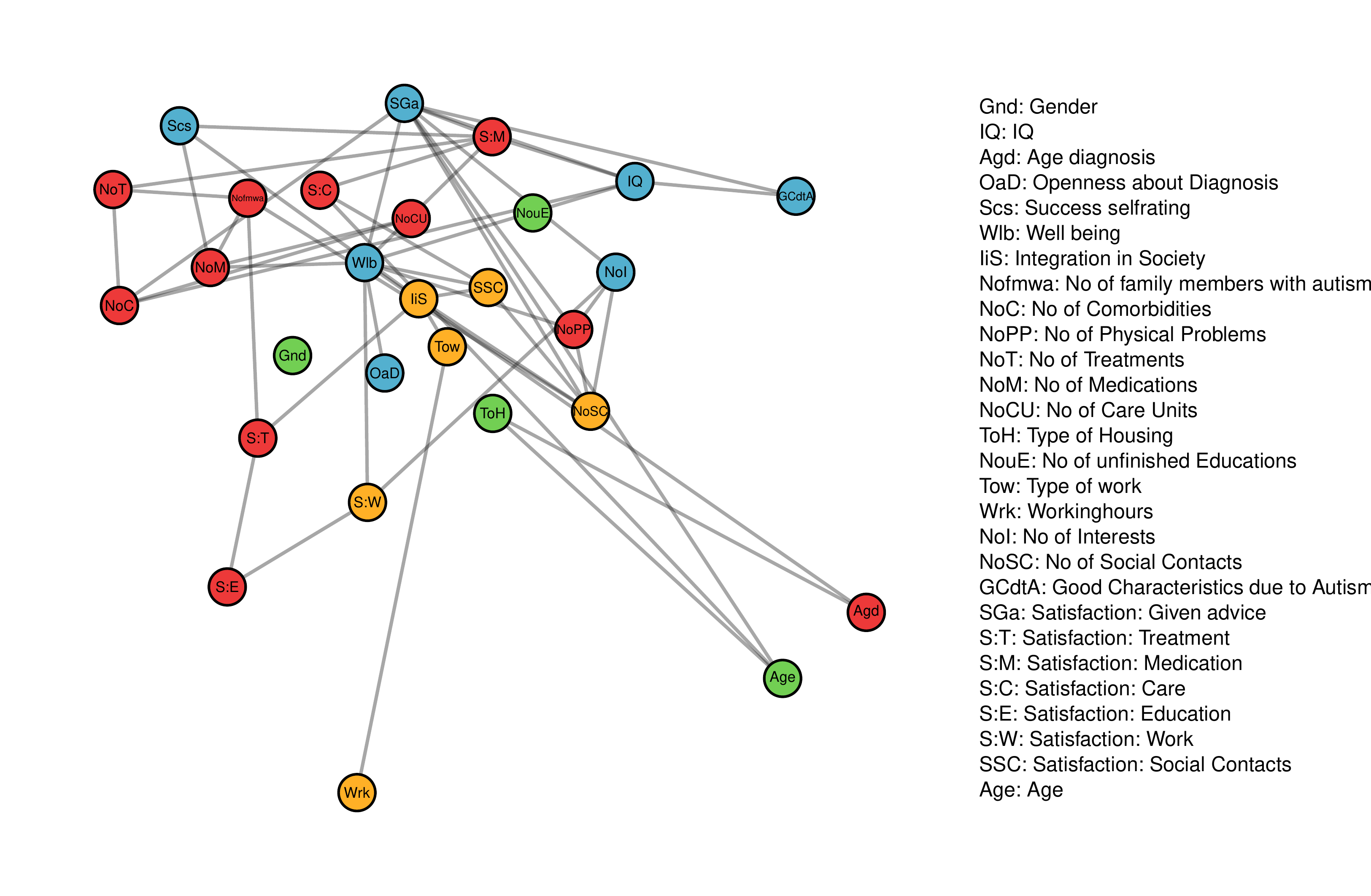}
(b) Gaussian Graphical Model

  \end{minipage} 
\caption{Comparing two graphical models: (a) all variables are modeled on their proper domain (b) all variables are modeled as Gaussians.}
\label{examplegraphs}
\end{figure}

We assumed that the maximal clique size in the true graph is two and we therefore augment second-order interactions $(k=2)$ to the covariance matrix and to combine parameters across regressions with the AND-rule. Different to the algorithm used for the simulations, we select the penalization-parameter $\lambda$ using the EBIC, because the cross-validation requirement that each category is present in each fold was not satisfied in this dataset. For the layout in Figure \ref{examplegraphs}, we used the force-directed algorithm of \cite{fruchterman_graph_1991}, that places nodes with many edges close to the center of the layout.

\begin{figure}
\centering
\begin{minipage}[b]{1\linewidth}
    \centering
	\includegraphics[width=1\textwidth]{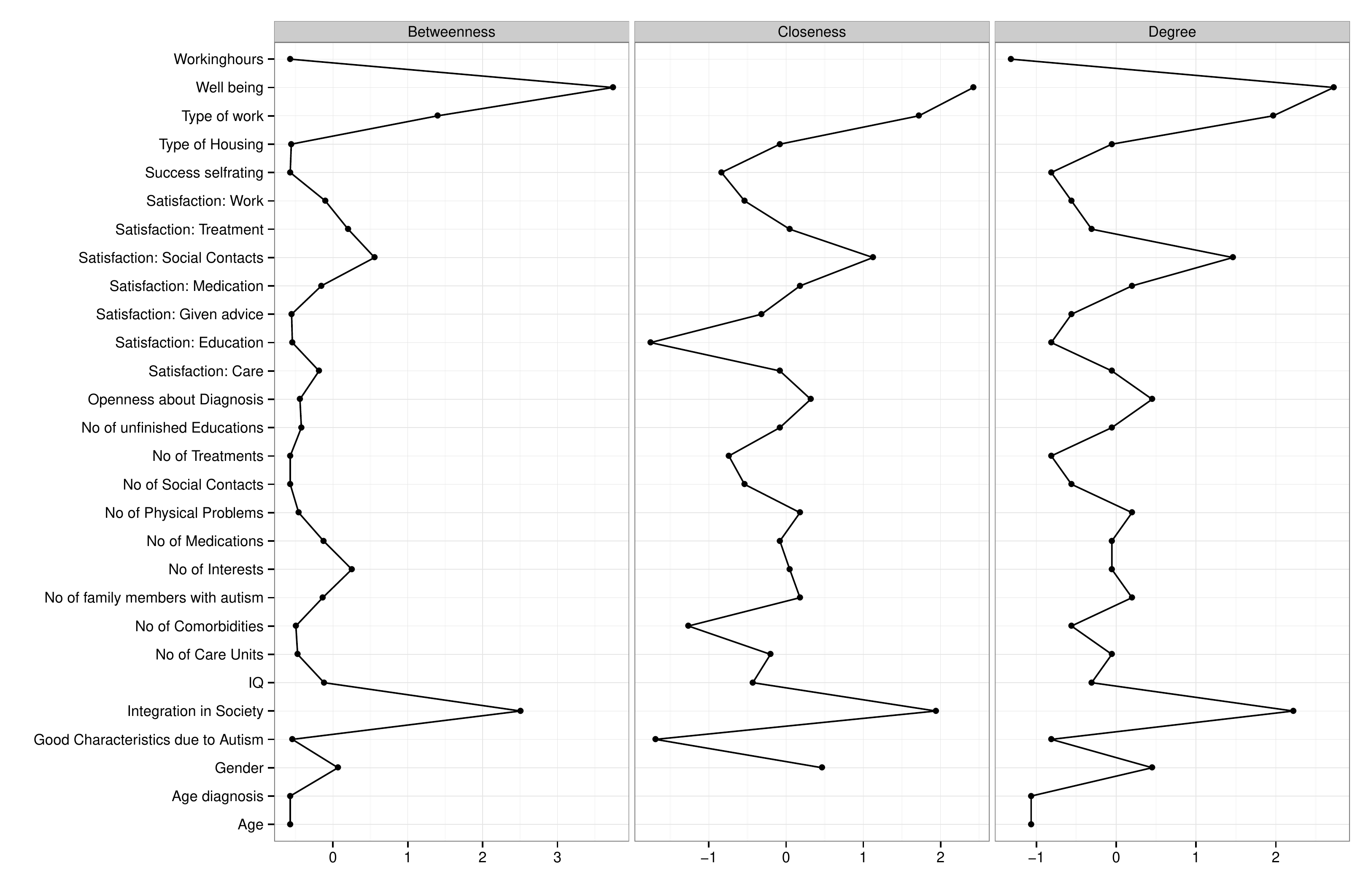}
(a) Mixed Graphical Model

  \end{minipage}
  \begin{minipage}[b]{1\linewidth}
    \centering
    \includegraphics[width=1\textwidth]{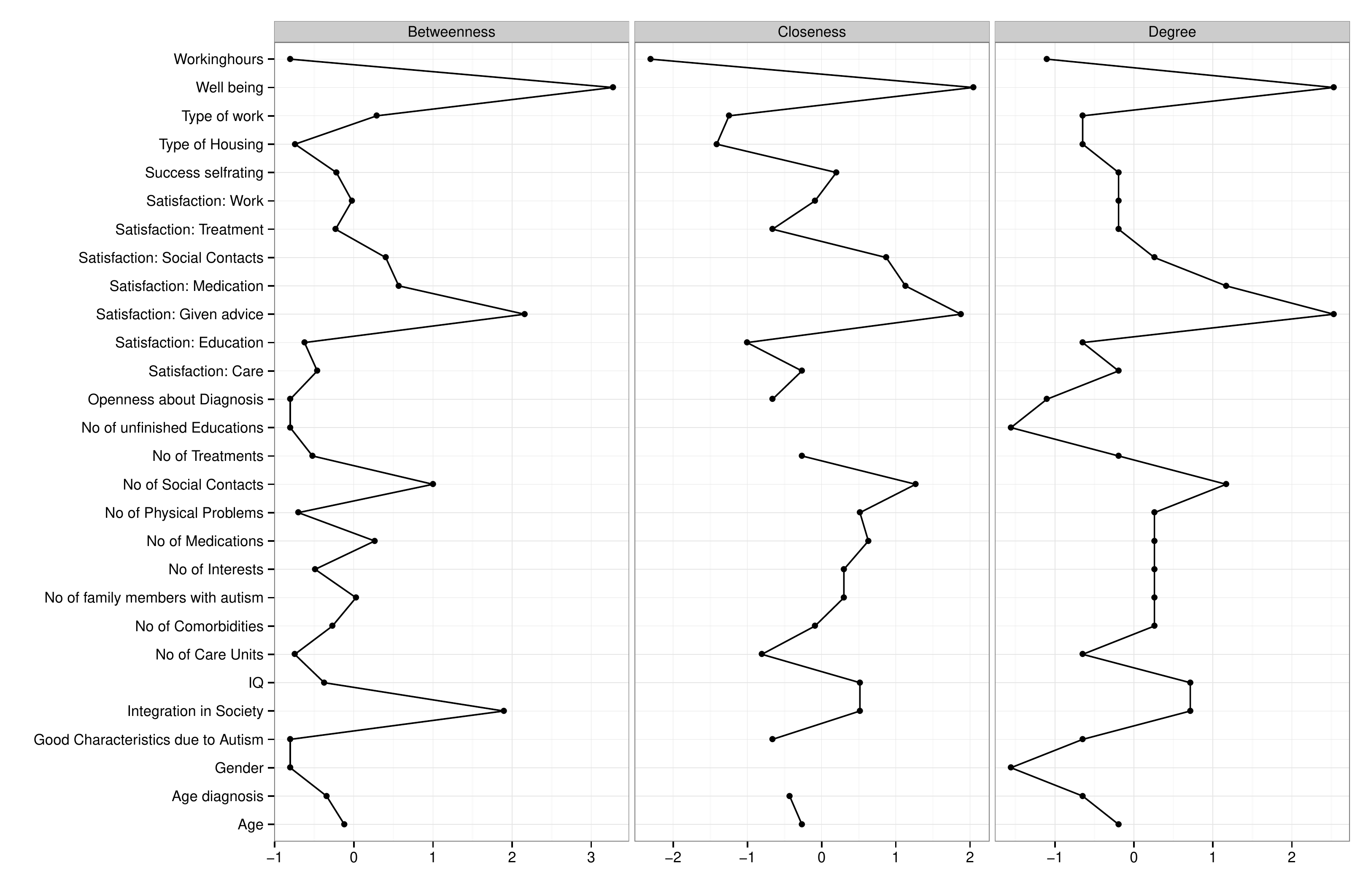}
(b) Gaussian Graphical Model

  \end{minipage} 
\caption{Comparing the centrality measures betweenness, closeness and degree of the mixed graphical model (a) and the Gaussian graphical model (b) in Figure \ref{examplegraphs}.}
\label{exgr_cp}
\end{figure}

In Figure \ref{examplegraphs} (a), we see that the different aspects 'Demographics', 'Psychological', 'Social environment', and 'Medical' are strongly interrelated, which highlights the need for an integrated analysis. On the level of single variables, we can visually identify the importance of single variables, for example we see that 'Well-being' is central in the graph and has unique associations with many other variables. We can make the analysis of single variables more explicit by computing centrality measures for each node. Centrality measures quantify the importance of a node in a network, where the exact interpretation of importance depends on the specific centrality measure. In Figure \ref{exgr_cp} (a) we report the standardized centrality measures betweenness, closeness and degree \cite[for details see][]{Opsahl_node_2010}. We see, for example, that 'Integration in Society' scores relatively high on closeness, degree and betweenness. This means 'Integration in Society' is relatively close to other nodes (closeness), has many connections (degree), and is often on the shortest path between any two nodes (betweenness).

We estimated the graph in Figure \ref{examplegraphs} (a) by modeling categorical variables as categorical variables, real-valued variables as Gaussian and count-valued variables as Poisson in a mixed graphical model. But does this actually give us a different graph compared with when we treat all variables as Gaussians? Figure \ref{examplegraphs} (b) shows the graph resulting from using the same estimation method as in (a), but treating all variables as Gaussians. We see that the resulting graph has less edges (density $ = .13$ vs. $.19$). However, method (b) is not simply more liberal. We see also edges that are present in (a) but not in (b) such as the edge between 'Type of housing' and 'Openness about Diagnosis'. Also by comparing the centrality plots in \ref{exgr_cp} (a) and (b) we see considerable differences. For example, 'Satisfaction with Social Contacts' is one of the most central nodes in the mixed graphical model, while in the Gaussian graphical model its centrality is average or below average. These substantial differences between the two graphs highlight the importance of modeling variables on their proper domain.

\section{Discussion}\label{concl}

In the present paper, we extended the generalized covariance method of \cite{loh_structure_2013} to the class of mixed distributions introduced by \cite{yang_mixed_2014}. We thereby provide an estimation method for the underlying graph structure of joint distributions over any combination of univariate members of the exponential family.

The peculiar simulation results of increasing and then decreasing precision with growing $n$ in conditions with $m=2,3$ and $d=1$ show that it is important to make sensible choices about how to combine the set of parameters involved in an interaction including a categorical variable into one dependence parameter. One solution to this problem might be using the group lasso \citep{yuan_model_2006, jacob_group_2009}, which induces sparse estimation on the group level. 

As mentioned in Section \ref{mixedMRF}, a limitation of the class of mixed distributions that in case they include two univariate distributions with infinite domain, they are not normalizable if \textit{neither} both distributions are infinite only from one side \textit{nor} the base measures are bounded with respect to the moments of the random variables. While this is an important theoretical problem that has to be addressed in future research, it does not limit the applicability of our method in most situations. Problems would only occur in the presence of extremely large values, which are rarely found in real-world data as most measurement scales are  (naturally) bounded.

As indicated in Section \ref{results}, we added additional noise after sampling to be able to perform 10-fold cross-validation to select an appropriate $\lambda_n$ parameter for the $\ell_1$-penalty in Algorithm \ref{alg_1}. The proportion of data points replaced by noise until the requirement for cross-validation was met is depicted in Figure \ref{noise_resamp} in Appendix \ref{app_1}. As can be seen in the figure, the proportion of added noise is considerable for the Binary-Gaussian and Binary-Exponential for small $\frac{n}{p}$ and in categorical graphs with $m=2,3$ and $\frac{n}{p}=1$. Therefore, in cases with a nonzero proportion of additionally added noise, the above simulation-results can interpreted as \textit{conservative} estimates of how well the method performs in practice.

In order to make our method available to researchers, we implemented our method as the R-package \textbf{mgm}, that is available from the Comprehensive R Archive Network (CRAN) at http://CRAN.r-project.org/.

Statistical analyses of multivariate data based on Markov random fields are becoming increasingly popular in many areas of science. Despite the fact that most datasets involve different types of variables, up until now, there was no principled method available to estimate the Markov random field underlying a joint distribution over different types of variables. We closed this methodological gap by providing a well-performing and easy to interpret method to estimate the underlying Markov random field of a joint distribution consisting of any combination of univariate exponential family members, which includes commonly used distributions such as Gaussian, Bernoulli, multinomial, Poisson, exponential, gamma, chi-squared and beta. We provided simulation results illustrating the performance of the method in realistic situations, illustrated our method with a network of different live aspects of individuals diagnosed with ASD and presented an implementation of our method as an R-package.


\bibliography{tbib}{}
\bibliographystyle{imsart-nameyear}

\newpage

\appendix

\section{Proofs of supporting Lemmas}\label{app_2}

\subsection{Proof of Lemma 1}\label{proof_l1}

The log-normalizing constant of the mixed joint distribution in (\ref{eq:mixed.joint.full}) can be written as

\begin{equation*}
\Phi(\theta) := \log \int_{x\in\mathcal{X}^p} 
\exp
\big \{ 
\sum_{C \in \mathcal{C}}  \theta_{C} \phi(x_{C}) 
\big\} \nu (dx),
\end{equation*}

\noindent
where $\nu$ is the count-measure, the Lebesgue-measure, or a combination of both.

To establish minimality, suppose $\sum_{C} a_{C} \phi(x_{C}) = b$ almost surely, where the coefficients $a_{C}$ are real-valued and $b$ is some constant. Plugging in $x$ such that $x_{s} = 0$ for all $s \in V$ and using the fact that all states (discrete random variables) and intervals (continuous random variables) have positive probability, we see that $b = 0$. Now assume that not all $a_{C}$ are equal to 0. Let $C'$ be a set of cliques such that $a_{C'} \not = 0$ and $|C'|$ is minimal. Plugging in $x$ such that $x_{C'} \not = 0$ and $x_{C \setminus C'} = 0$, we have

\begin{equation*}
\sum_{C} a_{C} \phi(x_{C}) = a_{C'}
\end{equation*}

by the minimality of $|C'|$. This contradicts the fact $a_{C'} \not = 0$. Hence, we conclude that the sufficient statistics $\phi(x_{C})$ are indeed linearly independent, implying that the class of mixed distributions as in (\ref{eq:mixed.joint.full}) is minimal.

\subsection{Corollary 2}\label{c_2}

Let $\textbf{pow}(\mathcal{A}) = \bigcup_{C \in \mathcal{A}} \text{pow}(C)$ be the union of all $2^{|C|}-1$ nonempty subsets of all cliques $C$.

\newtheorem{c2}{Corollary}
\begin{c1}
Take any triangulation $\widetilde{G}$of the graph $G$ and let $\mathcal{A}$ be the set of separator sets in $\widetilde{G}$. Then the inverse $\Gamma$ of the covariance matrix $\emph{cov}(\Psi(\phi(X);;V \cup \emph{\textbf{pow}}(\mathcal{A})))$ has the property that $\widetilde{\Gamma}(\{s\}, \{t\}) = 0$ whenever $(s,t) \notin \widetilde{E}$.
\end{c1}

A consequence of Corollary \ref{c_2} is that for all graphs with singleton separator sets (e.g. trees), the original covariance matrix $\text{cov}(\Psi(\phi(X), V))$ is graph structured. 

\textit{Proof.} The proof of Corollary \ref{c_2} follows directly from the proof of Theorem \ref{th_1}. We take the conjugate dual $\Phi^*(\mu)$ as in (\ref{triang_fact_dual}) but of the log normalization function $\Phi(\theta)$ corresponding to a model of the form (\ref{eq:mixed.joint.full}) including only terms for cliques $(V \cup \textbf{pow}(\mathcal{A}))$. Analogously to the proof of Theorem \ref{th_1} we take derivatives with respect to $\mu_s$ and $\mu_t$ and thereby all terms drop that do not contain both $\mu_s$ and $\mu_t$. Whenever $(s,t) \notin E$, all terms including both $\mu_s$ and $\mu_t$ are equal to zero, because $P_{\theta}(X)$ is Markov with respect to $G$. Therefore, whenever $(s,t) \notin E$, the partial derivative of $\Phi^*(\mu)$ with respect to  $\mu_s$ and $\mu_t$ is equal to zero. With equality (\ref{lw_key_eq}), this proves Corollary \ref{c_2}.

\section{Additional noise proportion in simulation}\label{app_1}

\begin{figure}[H]
\centering
\includegraphics[width=1\textwidth]{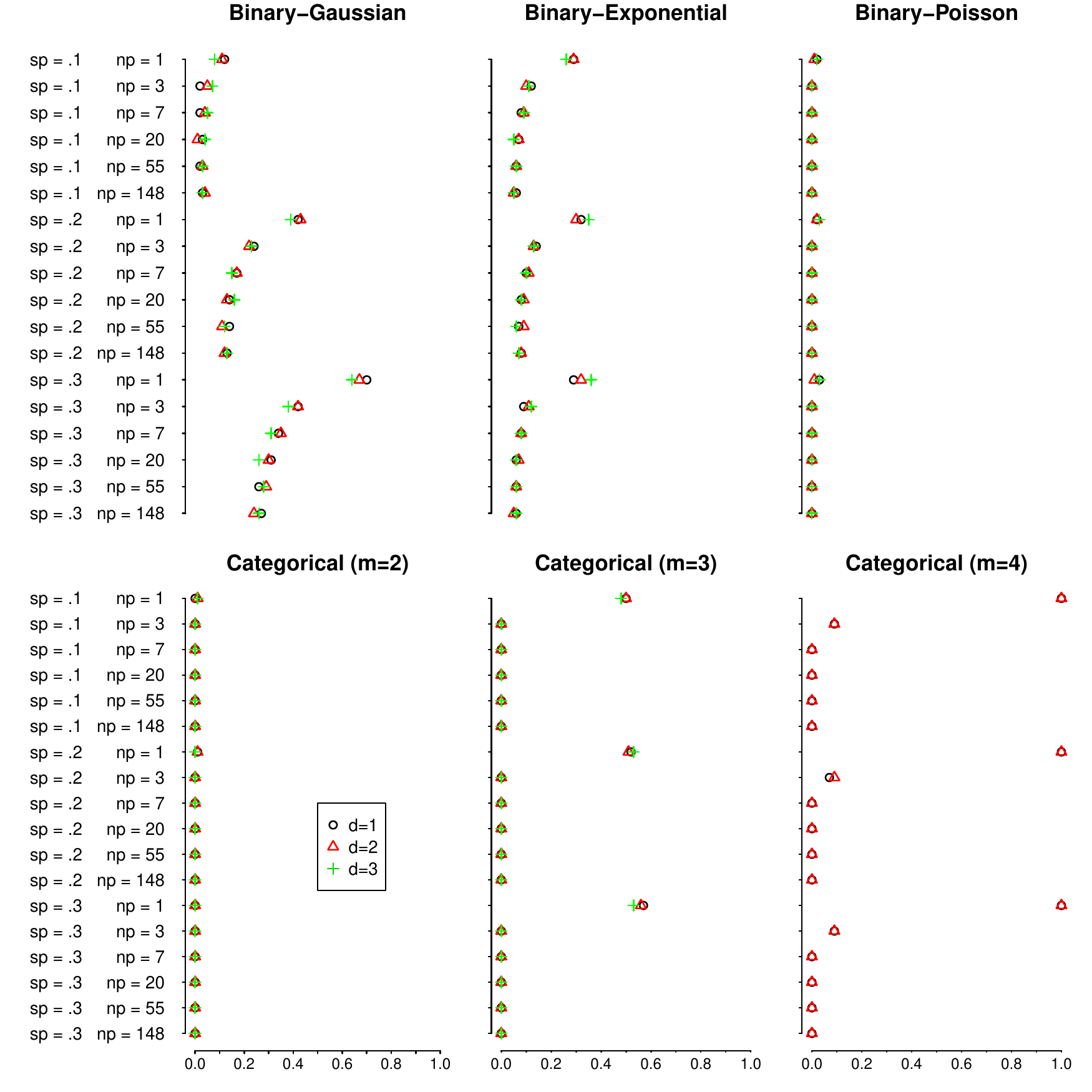}
\caption{Proportion of additional noise added to the data until the requirements for 10-fold cross-validation were met.}
\label{noise_resamp}
\end{figure}

Figure \ref{noise_resamp}  depicts the proportion of noise added after sampling the data until the requirements for 10-fold cross-validation were met. A proportion of $1$ means that all data points were replaced by noise.

\end{document}